\documentstyle[12pt]{article}
\topmargin - 0.5in
\oddsidemargin - 5mm
\begin{document}
\def\p{\phi}
\def\P{\Phi}
\def\a{\alpha}
\def\e{\varepsilon}
\def\be{\begin{equation}}
\def\ee{\end{equation}}
\def\l{\label}
\def\0{\setcounter{equation}{0}}
\def\h{\hat}
\def\b{\beta}
\def\S{\Sigma}
\def\C{\cite}
\def\r{\ref}
\def\ba{\begin{eqnarray}}
\def\ea{\end{eqnarray}}
\def\n{\nonumber}
\def\X{\Xi}
\def\x{\xi}
\def\la{\lambda}
\def\d{\delta}
\def\D{\Delta}
\def\s{\sigma}
\def\f{\frac}
\def\Ga{\Gamma}
\def\ga{\gamma}
\def\th{\theta}
\def\Th{\Theta}
\def\pa{\partial}
\def\o{\omega}
\def\O{\Omega}
\def\vae{\varepsilon}
\def\rar{\rightarrow}

\vskip 3cm
\begin{center}
{\Large \bf Perturbation theory in the invariant subspace}
\vskip 1cm
J.Manjavidze \\Institute of Physics, Tbilisi,
Rep. of Georgia \\

\end{center}
\vskip 1.5cm

\begin{abstract}\footnotesize
The unitary transformation of path-integral differential measure is described.
The main properties of perturbation theory in the phase space of action-angle,
energy-time variables are investigated. The measure in cylindrical coordinates
is derived also. The dependence of perturbation theory contributions from
global (topological) properties of corresponding phase spaces is shown.
\end{abstract}

\section{Introduction}\0

The main purpose of this article is to describe a quantum system with
nontrivial
phase-space topology. The approach will be illustrated, risking
to loose generality, by simplest quantum-mechanical examples of particle
motion in the potential hole $v(x)$ with one nondegenerate minimum at $x=0$.

There is a definite hope that offered below
formalism will be useful for quantization of nonlinear waves. Indeed, the
last problem was considered previously by many authors, e.g. \C{gold, kor},
introducing the convenient variables (collective coordinates) in order that
reduce the quantum soliton-like excitations problem to quantum-mechanical one.

Quantitatively the problem looks as follows. It is not difficult to
describe one-particles dynamics in the quasiclassical approximation
since corresponding equation for trajectory $x_c$ always
can be solved. But, beyond this approximation, to use the ordinary
WKB expansion of path integrals
one should solve the equation for Green function $G$:
\be
(\pa^2 + v''(x_c))_t G(t,t')=\d (t-t').
\l{2}\ee
Just eq.(\r{2}) offers a difficulty: it is impossible to find a strict
solution of this equation since $x_c =x_c(t)$ is the nontrivial function
and, therefore, $G(t,t')$ is not translationally invariant describing
propagation of a `particle' in the $time-dependent$ potential $v''(x_c)$.
One can hope to avoid this difficulty introducing the convenient dynamical
variables. Demonstration of the way as this program can
be realized is the aim of this article.

The main formal difficulty, e.g. \C{mar}, of this program consisting in
transformation of the path-integral measure was solved in \C{yad}. It was
shown in this paper that the phase-space $(x,p)$ path integrals differential
measure $DM(x,p)$ of $probability-like$ quantities $R \sim |A|^2$, where $A$
is the amplitude, is $\d$-like:
\be
DM(x,p)=\prod_t dx(t)dp(t)\d (\dot{x} + \frac{\pa H_j}{\pa p})
\d (\dot{p} - \frac{\pa H_j}{\pa x}),
\l{3}\ee
where the Hamiltonian
$$
H_j =\frac{1}{2}p^2 + v(x) -jx
$$
includes the energy of quantum fluctuations $jx$, with the provoking
quantum excitations force $j=j(t)$. The dynamical equilibrium between
ordinary mechanical forces (kinetic $\dot{p}(t)$ plus potential $v'(x)$) and
quantum force ($j(t)$) fixed by $\d$-like measure (\r{3}) allows
to perform an arbitrary transformation of quantum measure caused by
transformation of classical forces, i.e. of $x$ and $p$.

We will use this property introducing the `motion' on the cotangent bundle
($\th ,h$), where $h$ is the bundles parameter and $\th$ is the coordinate
on it. For definiteness, let $h$ be the conserved on the $classical$
trajectory energy and $\th$ is the conjugate to $h$ `time'. (The
transformation to action-angle variables will be described also.)
The mapping
$(x,p) \rightarrow (\th ,h)$ is canonical and the corresponding
equations of motion on the cotangent bundle should have the form:
\ba
\dot{h}=-\frac{\pa H_c}{\pa \th}=+j\frac{\pa x_c (\th ,h)}{\pa \th},
\n \\
\dot{\th}=+\frac{\pa H_c}{\pa h}=1-j\frac{\pa x_c (\th ,h)}{\pa h},
\l{4}\ea
where
$$
H_c =h -jx_c (\th ,h)
$$
is the transformed Hamiltonian and $x_c (\th ,h)$ is the classical trajectory
in the $(\th ,h)$ terms. The Green function of the eq.(\r{4}) $g(t,t')$ is
translationally invariant since classically (at $j=0$) the cotangent
bundle is the time-independent manifold.

Above example shows that the mapping is constructive iff the bundle
parameters are generators of (sub)group violated by $x_c$. Corresponding
phase space is the invariant subspace \C{smale}. For some problems the
mapping on the group manifolds is useful, see, e.g., \C{dow}. We will
demonstrate the noncanonical transformation to cylindrical coordinates.

The perturbation theory with measure (\r{3}) on the cotangent bundles
has unusual properties \C{untr, ang} and in Sec.3 few proposition
concerning perturbation theory on the invariant subspaces will be offered.
The main of them demonstrate that each order of perturbation theory can
be reduced to the total derivative over global coordinate of the invariant
subspace. The application of derived formalism in quantum mechanics and
field theories will be published later. For completeness
in the following Sec.2 the origin of approach \C{yad} will be shown.

\section{Unitarity condition}\0

It is known that the transformations of the path-integral measure deform
perturbation theory contributions uncontrollably because of the stochastic
nature of quantum trajectories \C{mar}. Purpose of this Section is to show
how the $S$-matrix unitarity condition can be introduced into the path-
integral formalism to find measure (\r{3}) (preliminaries were given in
\C {yad}).

The unitarity condition for the  $S$-matrix $SS^+ =S^+ S=I$ presents the
infinite set of nonlinear equalities:
\be
iA A^* =A - A^*,
\l{2'}\ee
where $A$ is the amplitude, $S=I+iA$. Expressing the amplitude by the
path integral one can see that the l.h.s. of this equality offers the
double integral and, at the same time, the r.h.s. is the linear combination
of integrals. Let us consider what this linearization of product $AA^*$ gives.

Using the spectral representation of one-particle amplitude:
\be
A(x_1 ,x_2 ;E)=\sum_{n}\frac{\Psi^*_{n} (x_2)\Psi_n (x_1)}{E-E_n +i\vae},
~~~\vae \rar +0,
\l{3'}\ee
let us calculate
\be
R(E)=\int dx_1 dx_2 A(x_1 ,x_2 ;E)A^* (x_1 ,x_2 ;E).
\l{4'}
\ee
The integration over end points $x_1$ and  $x_2$ is performed for sake
of  simplicity only. Using ortho-normalizability of the wave functions
$\Psi_{n} (x)$ we find that
\be
R(E)=\sum_{n}|\frac{1}{E-E_n+i\vae}|^2 =\frac{\pi}{\vae}\sum_{n}\d (E-E_{n}).
\l{5'}\ee
Certainly, the last equality is nothing new but it is important to note that
$R(E) \equiv 0$ for all $E\neq E_n$, i.e. that all unnecessary contributions
with $E\neq E_n$ were canceled by difference in the r.h.s. of eq.(\ref{2'}).
We will put this phenomena in the basis of the approach.

We will build the perturbation theory for $R(E)$ using the path-integral
definition of amplitudes \C{yad}. It leads to loss of some information since
the amplitudes can be restored in such formulation with phase accuracy only
\footnote{One can use the dispersion relations to reconstruct the amplitude
from $R(E)$}.
Yet, it is sufficient for calculation of the energy spectrum. We would
consider this quantity to demonstrate following statement:

{\it S1. The unitarity condition unambiguously determines contributions
in the  path integrals for $R(E)$.}

This statement looks like a tautology since $\exp\{iS(x)\}$, where $S(x)$
is the quantum-mechanical action, is the unitary operator which shifts a
system along the trajectory\footnote{It is well known that this unitary
transformation is the analogy of tangent transformations of classical
mechanics \C{fok}.}. I.e. the unitarity is already fixed in the path-
integrals. But the general path-integral solution contains unnecessary
degrees of freedom (unobservable states with $E\neq E_n$ in our example).
We want define the quantum measure $DM$ in such a way that the condition
of absence of unnecessary contributions in the final (measurable) result
be loaded from the very beginning. Just in this sense the unitarity
looks like the necessary and sufficient condition unambiguously determining
the complete set of contributions. Solution is simple: one should search, as
it follows from(\r{5'}), the linear path-integral representation for $R(E)$
to introduce this condition into the formalism.

Indeed, to see the integral form of our approach, let us use
the proper-time representation:
\be
A(x_1 ,x_2 ;E)=\sum_{n}
\Psi_{n} (x_1)\Psi^{*}_{n} (x_2)i
\int^{\infty}_{0}dTe^{i(E-E_{n}+i\varepsilon)T}
\l{6'}
\ee
and insert it into (\ref{4'}):
\be
R(E)=\sum_{n}
\int^{\infty}_{0}
dT_{+}dT_{-} e^{-(T_{+}+T_{-})\varepsilon}
e^{i(E-E_{n})(T_{+}-T_{-})}.
\l{7'}
\ee
We  will introduce new time variables instead of $T_{\pm}$:
\be
T_{\pm}=T\pm\tau,
\l{8'}
\ee
where, it follows  from Jacobian of the transformation,
$|\tau|\leq T,\;\;\;0\leq T\leq \infty$. But
we can put $|\tau|\leq\infty$ since $T\sim1/\varepsilon\rightarrow\infty$
is  essential in integral over $T$. In result,
\be
R(E)=2\pi\sum_{n}\int^{\infty}_{0}
dT e^{-2\varepsilon T}
\int^{+\infty}_{-\infty}\frac{d\tau}{\pi}
e^{2i(E-E_{n})\tau}.
\l{9'}
\ee
In the last integral all contributions with $E\neq E_{n}$ are
canceled.
Note that the  product of amplitudes $AA^*$ was `linearized' after
introduction of `virtual' time $\tau =(T_{+}-T_{-})/2$. The
physical meaning of such variables will be discussed, see also \C{fok}.

We will consider following path-integral:
\be
A(x_1 ,x_2 ;E)=i\int^{\infty}_{0}dT e^{iET}\int Dx e^{iS_{C_+}(x)}
\delta (x_1 -x(0))\delta (x_2 -x(T)),
\l{3.1}\ee
where
$$
C_+ : t \rar t+i\vae, \vae \rar +0,
$$
is the Mills complex time contour \C{mill}. Calculating the probability
to find a particle with energy $E$ ($Im~E$ will not be mentioned for
sake of simplicity) we have:
\ba
R(E)=\int dx_1 dx_2 |A|^2 =\int^{\infty}_{0} dT_+ dT_- e^{iE(T_+ -T_-)}
\int D_{C_+}x_+  D_{C_-}x_-\times
\n \\
\delta (x_+ (0) -x_- (0))\delta (x_+ (T_+ ) -x_- (T_- ))
e^{iS_{C_+ (T_+ )}(x_+ ) - iS_{C_- (T_- )}(x_- )},
\l{3.2}\ea
where $C_- (T)=C^{*}_{+}(T)$. Note that the total action in (\ref{3.2})
$S_{C_+ (T_+ )}(x_+ ) - S_{C_- (T_- )}(x_- )$ describes the closed-path
motion by definition.

New time variables $T$ and $\tau$ will be used: $T_{\pm}=T\pm\tau$. If
$Im~E\rar +0$ then $T$ and $\tau$ can be considered as the
independent variables: $0\leq T \leq \infty$,$-\infty\leq\tau\leq\infty$.
We will introduce also the mean trajectory $x(t)=(x_+(t) +x_-(t))/2$ and
the deviation $e(t)$ from it: $x_{\pm}(t)=x(t)\pm e(t)$. Note that one can
do surely this linear transformations in the path integrals if the space
is flat. We will consider $e(t)$ and $\tau$ as the fluctuating, virtual,
quantities and calculate the integrals over them perturbatively. In zero
order over $e$ and $\tau$, i.e. in the quasiclassical approximation, $x$
is the classical path and $T$ is the total time of classical motion.

The boundary conditions (see (\r{3.2})) should fix the closed-path motion
and therefore we have the boundary conditions for $e(t)$ only:
\be
e(0)=e(T)=0.
\l{3.6}\ee
Note the uniqueness of this solution if the integral over $\tau$ is calculated
perturbatively.

Extracting the linear over $e$ and $\tau$ terms from the closed-path action
$S_{C_+ (T_+ )}(x_+ ) - S_{C_- (T_- )}(x_- )$ and expanding over $e$ and
$\tau$ the remainder terms:
\be
-\tilde {H}_T (x;\tau )=S_{C_+ (T+\tau)}(x) - S_{C_- (T-\tau )}(x)+
2\tau H_T (x),
\l{3.7}\ee
where $H_T(x)$ is the Hamiltonian at the time moment $T$, and
\be
-V_T (x,e)=S_{C_+ (T)}(x+e)-S_{C_- (T)}(x-e)+2Re\int_{C^{(+)}} dt e(\ddot{x}+
v'(x))
\l{3.8}\ee
we find that
\be
R(E)=2\pi \int^{\infty}_{0}dTe^{-i\h{K}(\o ,\tau;j,e)}
\int DM(x) e^{-i\tilde{H}_T (x;\tau)-iV_T (x,e)}.
\l{3.10}\ee

The  expansion over differential operators:
\be
\h{K}(\o ,\tau;j,e)=
\frac{1}{2}(\frac{\partial}{\partial\o}\frac{\partial}{\partial \tau}
+Re\int_{C_+(T)}dt\frac{\delta}{\delta j(t)}\frac{\delta}{\delta e(t)})
\l{3.11}\ee
will generate the perturbation series. We propose that it exist in Borel
sense.

In (\r{3.10}) the functional measure
\be
DM(x)=\delta (E+\o -H_T (x))\prod_t dx(t) \delta (\ddot{x}+v'(x)-j)
\l{3.12}\ee
unambiguously defines the $complete$ set of contributions in the path
integral. The
functional $\delta$-function is defined as follows:
\ba
\prod_t \delta (\ddot{x}+v'(x)-j)=(2\pi )^2 \int \prod_{t}\frac{de(t)}{\pi}
\delta(e(0))\delta(e(T))e^{-2iRe\int_{C_+}dt e(\ddot{x}+v'(x)-j)}=
\n \\
\prod_{t\in C_+ (T)}\delta (Re (\ddot{x}+v'(x)-j))\delta(Im (\ddot{x}+v'(x)-j))
\l{*}\ea

The physical meaning of this $\d$-function is following. We can consider
$(\ddot{x}+v'(x)-j)$ as the total force and $e(t)$ as the virtual
deviation from true trajectory $x(t)$. In classical mechanics the virtual
work must be equal to zero: $(\ddot{x}+v'(x)-j)e(t)=0$ (d'Alembert) \C{arn}
since the motion is time reversible. From this evident dynamical principle
one can find the `classical' equation of motion:
\be
\ddot{x}+v'(x)=j,
\l{3.13}\ee
since $e(t)$ is arbitrary.

Generally, in quantum theories the virtual work is not equal to zero, i.e.
the quantum motion is not time reversible since the quantum corrections can
shift the energy levels. But
integration over $e(t)$, with boundary conditions (\ref{3.6}), leads to
the same result. So, in quantum theories the unitarity condition (i.e. the
destructive interference among two exponents in product $AA^*$
\C{fok}) play the same role as the d'Alembert's variational principle in
classical mechanics. We can conclude, the unitarity condition as the
dynamical principle establish the $local$ equilibrium between classical
(r.h.s. of (\r{3.13})) and quantum (l.h.s. of (\r{3.13})) forces. $\bullet$

\section{Perturbation theory}\0

Now let us consider representation (\r{3.10}). It is not hard to show that

{\it S2. Eq.(\r{3.10}) restores the perturbation theory of stationary phase
method}.

For this purpose it is enough to consider the ordinary integral:
\be
A(a,b)=\int^{+\infty}_{-\infty}\frac{dx}{(2\pi)^{1/2}}e^{i(\frac{1}{2}ax^2
+\frac{1}{3}bx^3)},
\l{3.14}
\ee
with $Im~a \rar +0$ and $b>0$. Computing the `probability' $R=|A|^2$ we find:
\be
R(a,b)=e^{\frac{1}{2i}\hat{j}\hat{e}}\int^{+\infty}_{-\infty} dx
e^{-2(x^2 +e^2 )Im~a}e^{2i\frac{b}{3}e^3}\delta (Re~ax +bx^2+j).
\l{3.1'}\ee

Performing the trivial transformation $e\rar ie$, $\hat{e}\rar -i\hat{e}$ of
auxiliary variable we find at the limit $Im~a=0$ that the contribution of
$x=0$ extremum (minimum) gives expression:
\be
R(a,b)=\frac{1}{a}e^{-\frac{1}{2}\hat{j}\hat{e}}(1-4bj/a^2)^{-1/2}
e^{2\frac{b}{3}e^3}
\l{3.15}\ee
and the expansion of operator exponent gives the asymptotic  series:
\be
R(a,b)=\frac{1}{a}\sum^{\infty}_{n=0}(-1)^{n}\frac{(6n-1)!!}{n!}
(\frac{2b^4}{3a^6})^n,~~~(-1)!!=0!!=1.
\l{3.16}\ee
This series is convergent in Borel sense.

Eq.(\r{3.1'}) can be considered as the definition of integral (\r{3.14}).
By this reason one may put $Im~a =0$ from the very beginning. We will use
this property.

Let us calculate now $R$ using the stationary phase method. Contribution
from the minimum $x=0$ gives $(Im~a=0)$:
$$
A(a,b)=e^{-i\hat{j}\hat{x}}e^{-\frac{i}{2a}j^2}e^{i\frac{b}{3}x^3}
(\frac{i}{a})^{1/2}.
$$
The corresponding `probability' is
\be
R(a,b)=\frac{1}{a}e^{-\frac{1}{2}\hat{j}\hat{e}}e^{2\frac{b}{3}e^3}
e^{\frac{2b}{a^2}ej^2}
\l{3.17}\ee
This expression does not coincide with (\ref{3.15}) but it leads to
the same asymptotic series (\r{3.16}). $\bullet$

The solution $x_j (t)$ of eq.(\r{3.13}) we would search expanding it over
$j(t)$: $$x_j (t)=x_c (t)+\int dt_1 G(t,t_1 )j(t_1 )+...$$
This is sufficient since $j(t)$ is  the auxiliary variable. In this
decomposition $x_c (t)$ is the strict solution of unperturbated equation
$\ddot{x}+v'(x)=0$ and $G(t,t')$ must obey eq.(\r{2}). Note that the
functional $\d$-function in (\ref{3.12})
does not contain the end-point values of time $t=0$ and $t=T$. This means
that the initial conditions to the eq.(\ref{3.13}) are not fixed and the
integration over them is assumed because of the definition of $R$.

The $\d$-likeness of measure allows to conclude:

{\it S3. All strict regular solutions (including trivial) of classical
(unperturbated by $j$) equation(s) of motion must be taken into account.}

We must consider only `strict' solutions because of strict cancellation of
needless contributions. The $\d$-likeness of measure means that the
probability $R(E)$ should contain $sum$ over all discussed solutions. This
is the main distinction of our unitary method of quantization from
stationary phase method: even having few solutions there is not interference
terms in the sum over them in $R$.

Note that the interference terms are absent independently from solutions
`nearness' in the functional space. This reflects the orthogonality of
Hilbert spaces builded on the various $x_c$ \C{gold} and is the consequence
of unitarity condition.

Summation over all solutions of classical equation of motion means necessity
to take into account all topologically-equivalent orbits $x_c$ also. This
naturally gives integration over zero-mode degrees of freedom. The
corresponding measure will be defined by mapping on the cotangent bundle
(see $S4$).

It is evident that in the sum over contributions of various $x_c$ we must
leave largest, i.e. with maximal number of zero modes. This selection rule
\C{yad} is equivalent of definition of the vacuum in the canonical
quantization scheme.

The solutions must be regular since the singular $x_c$ gives zero
contribution on $\d$-like measure. $\bullet$

It is evident that

{\it S4. The measure (\r{3.12}) admits the canonical transformations}.

This evidently follows from $\d$-likeness of measure. In the phase space
we have:
\be
DM(x,p)=\delta (E+\o -H_T (x))\prod_{t}dx dp \delta(\dot{x}-\frac{\partial H_j}{\partial p})
\delta(\dot{p}+\frac{\partial H_j}{\partial x}),
\l{3.18}\ee
where
\be
H_{j}=\frac{1}{2}p^2 +v(x)-jx
\l{3.19}\ee
is the total Hamiltonian which is time dependent through $j(t)$.

Instead of pare $(x,p)$ we can introduce new pare $(\theta ,h)$ inserting
\be
1=\int D\theta Dh\prod_{t}\delta(h-\frac{1}{2}p^2 -v(x))
\delta(\theta -\int^{x}dx (2(h-v(x)))^{-1/2}).
\l{3.20}
\ee
It is important that both differential measures in (\r{3.20}) and (\r{3.18})
are $\delta$-like. This allows to change the order of integration surely and
firstly integrate over $(x,p)$. Calculating result one can use $\d$-functions
of (\r{3.18}). In this case $\d$-functions of (\r{3.20}) will define the
constraints. But if we will use $\d$-functions of (\r{3.20}) the mapping
$(x,p)\rightarrow (\th ,h)$ is performed. We conclude that our transformation
takes into account the constraints since both ways must give the same
result.

We find:
\be
DM(\th ,h)=\delta (E+\o -h(T))\prod_{t}\delta(\dot{\theta}-\frac{\partial H_c}{\partial h})
\delta(\dot{h}+\frac{\partial H_{c}}{\partial \theta}),
\l{3.21}\ee
since considered transformation is canonical, $\{h(x,p),\theta (x,p)\}=1$,
where
\be
H_c =h-jx_c (h,\theta)
\l{3.22}\ee
is the transformed Hamiltonian and $x_c (\theta,h)$ is the classical
trajectory parametrized by $h$ and $\theta$\footnote{I.e., the argument of
second $\d$-function in (\r{3.20}) becomes equivalent to zero at $x=x_c$.}.
$\bullet$

So, on the cotangent bundle we must solve following equations of motion:
\be
\dot{h}=j\frac{\partial x_c}{\partial \theta},
~~~\dot{\theta}=1-j\frac{\partial x_c}{\partial h},
\l{3.23}
\ee
which have a simple structure:

{\it S5. The Green function on the cotangent bundle is simple
$\Th$-function}.

Indeed, expanding solutions of eqs.(\r{3.23}) over $j$ in the zero order we
have $\th_0 =t_0 +t$ and $h_0 =const$. In the first order we have equation
for Green function $g(t,t')$:
\be
\dot{g}(t,t')=\d (t-t').
\l{3.24a}\ee
Noting that the $S$-matrix problem with definite boundary conditions
$x(0)=x_1$ and $x(T)=x_2$ is solved and $\d$-like measure was arise
in result of destructive interference between expanding and converging
waves in $AA^*$, i.e. wishing to describe the quantum motion from $x(0)
=x_1$ to $x(T)=x_2$ we without fail introduce the time `irrevercibility':
\be
g(t,t')=\Th (t-t'),
\l{3.24}\ee
in opposite to causal particles propagator $G(t,t')$, which contains both
retarded and advanced parts. But, as will be seen below, see $S10$, the
perturbation theory with Green function (\r{3.24}) is time reversible.
Note also, that the solution (\r{3.24}) is the unique and is the direct
consequence of usual in the quantum theories $i\e$-prescription.

The uncertainty is
contained in the boundary value $g(0)$. We will see that $g(0)=0$
excludes some quantum corrections. By this reason one should consider
$g(0)\neq 0$. We will assume that
\be
g(0)=1
\l{3.24b}\ee
since this boundary condition to eq.(\r{3.24a}) is natural for local theories.
We will use also following formal equalities:
\be
g(t,t')g(t',t)=0,~~~1=g(t,t')+g(t',t)
\l{3.24c}\ee
since $g(t,t')$ is introduced for definition of boundaries of time integrals.
$\bullet$

It is important to note that $Im~g(t)=0$ on the real time axis. This
allows to conclude that

{\it S6. The perturbation theory on the ($h,\th$) bundle can be
constructed on the real-time axis.}

Indeed, the $i\e$-prescription is not necessary since, as was mentioned
above in $S2$, the $\d$-functional
measure defines a complete set of contributions. But for more confidence
one may introduce the $i\e$-prescription and, extracting the $\d$-function in
the measure, one can put $\e =0$ if the contributions are regular at this
limit.

One can point out the examples when $\e =0$ is the singular point.

(i) The Green function G(t,t') is singular at $\e =0$. The
$i\e$-prescription introduces the wave damping in this case.

(ii) The terms of perturbation theory are singular at $\e =0$ even if the
Green functions are regular. This singularities are connected with light-cone
singularities of the real-time theories.

(iii) There is the tunneling phenomena. The $i\e$-prescription is necessary
to define a theory in the turning points (it is the usual WKB prescription).

(iv) The extremum of the action is unstable (is the maximum). In this case
the $\e =0$ limit is absent for corresponding contribution and one should
omit it (as in example considered in S2).

Considered in this paper examples allows the shift on the real-time axis.
$\bullet$

Note now that $\partial x_c/\partial \theta$ and $\partial x_c/\partial h$
in the r.h.s. of (\r{3.23})  can be  considered as the sources. This allows
to offer the statement:

{\it S7. The mapping on the cotangent bundle splits `Lagrange' quantum force
$j$ on a set of quantum forces individual to each independent degree of
freedom, i.e. to each independent local coordinate of the cotangent manifold}.

To show the splitting mechanism let us consider the action of the
perturbation-generating operators:
\ba
e^{-i\f{1}{2}Re\int_{C_+} dt \hat{j}(t)\hat{e}(t) }e^{-iV_T (x_c,e)}
\prod_{t} \d (\dot{h} -j\frac{\pa x_c}{\pa \th})
\d (\dot{\th} -1 +j\frac{\pa x_c}{\pa h}=
\n \\
=\int De_h De_{\th}e^{2iRe\int_{C_+}dt (e_h \dot{h}+e_{\th}(\dot{\th}-1))}
e^{-iV_T (x_c,e_c )},
\l{22}\ea
where
\be
e_c =e_h \frac{\pa x_c}{\pa \th} -e_{\th} \frac{\pa x_c}{\pa h}.
\l{23}\ee
The integrals  over $(e_h ,e_{\th})$ will be calculated perturbatively:
\be
e^{-iV_T (x_c,e_c )}=\sum^{\infty}_{n_h ,n_{\th} =0}\frac{1}{n_h !n_{\th} !}
\int \prod^{n_h }_{k=1}(dt_k e_h (t_k))\prod^{n_{\th} }_{k=1}(dt'_k e_{\th} (t'_k))
P_{n_h ,n_{\th}} (x_c ,t_1 ,...,t_{n_h},t'_1,...,t_{n_{\th}}),
\l{24}
\ee
where
\be
P_{n_h ,n_{\th}} (x_c ,t_1 ,...,t_{n_h},t'_1,...,t_{n_{\th}})=
\prod^{n_h }_{k=1} \hat{e}'_h (t_k)
\prod^{n_{\th} }_{k=1} \hat{e}'_{\th} (t'_k)e^{-iV_T (x_c,e'_c )}
\l{25}
\ee
with $e'_c \equiv e_c (e'_h ,e'_{\th} )$ and the derivatives in this
equality are calculated at $e'_h =0$, $e'_{\th}=0$. At the same time,
\be
\prod^{n_h }_{k=1} e_h (t_k)\prod^{n_{\th} }_{k=1} e_{\th} (t'_k)=
\prod^{n_h }_{k=1} (i\hat{j}_h (t_k))\prod^{n_{\th} }_{k=1}
(i\hat{j}_{\th} (t'_k))
e^{-2iRe\int_{C_+} dt (j_h (t)e_h (t)+j_{\th}(t)e_{\th}(t))}.
\l{26}\ee
The limit $(j_h , j_{\th})=0$ is assumed. Inserting (\ref{25}),
(\ref{26}) into (\ref{22}) we find new representation for $R(E)$:
\ba
R(E)=2\pi \int^{\infty}_{0} dT\exp\{\frac{1}{2i}(\hat{\omega}\hat{\tau}+
Re\int_{C_+} dt (\hat{j}_h (t)\hat{e}_h (t)+
\hat{j}_{\th} (t)\hat{e}_{\th} (t)))\} \times
\n \\
\int Dh D\th e^{-i\tilde{H}(x_c ;\tau )-iV_T (x_c ,e_c )}
\delta (E+ \omega -h(T))\prod_{t} \delta (\dot{h} -j_h )\delta (\dot{\th} -1 -
j_{\th}),
\l{27}\ea
in which the `energy' and the `time' quantum degrees of freedom are
splitting.

Therefore, splitting $j \rightarrow (j_h ,j_{\th})$
we must change $e \rightarrow e_c$, where $e_c$ carry the simplectic
structure of Hamilton's equations of motion, see (\r{23}), i.e. $e_c$
is the invariant of canonical transformations. This quantity describes the
flow $\d_{h}x_c \bigwedge \d_{\th}p_c$ generated by quantum perturbations
through the bundles elementary cell.

Hiding the $x_c (t)$ dependence in $e_c$ we had solve the problem of the
functional determinants and simplify the equation of motion as much as
possible:
\be
DM(h,\th )=\d (E+\o -h(T))\prod_{t} dh(t) d\th(t) \d (\dot{h}(t))
\d (\dot{\theta}(t)-1)
\l{3.26}\ee
and the perturbations generating operator
\be
\h{K}=\frac{1}{2}(\hat{\o}\hat{\tau}+
\int^T_0 dt_1 dt_2 \Theta (t_1 -t_2 ) (\hat{e}_h (t_1 )\hat{h}'(t_2 )+
\hat{e}_{\theta}(t_1 )\hat{\theta}' (t_2 )).
\l{3.27}\ee
In $V_T (x_c,e_c)$ we must change $h \rightarrow (h + h')$ and
$\th \rightarrow (\th +\th')$. $\bullet$

To describe quantum dynamics on the group manifolds we can consider also
the coordinate transformations. For instance, the two dimensional model with
potential $v=v((x^{2}_{1}+x^{2}_{2})^{1/2})$ should be considered in the
cylindrical coordinates $x_1 =r \cos\phi$, $x_2 =r \sin\phi$. Note that this
transformation is not canonical. In result we will see that

{\it S8. The transformed measure can not be deduced from direct
transformations of path integral (\r{3.1})}.

Indeed, the measure in the cylindrical coordinates
\be
D^{(2)}M(r, \phi ) =\delta (E+\o -H_T (x))\prod_t dr d\phi r^2 (t)
\delta (\ddot{r}-\dot{\phi}^2 r+v'(r)-j_r)\delta(\partial_t (\dot{\phi}r^2 )-rj_{\phi}),
\l{3.28}\ee
where $v'(r)=\partial v(r)/\partial r$ and $j_r$, $j_{\phi}$ are the
components of $\vec{j}$ in the cylindrical coordinates.

The perturbation generating operator has the form:
\be
\h{K}=\frac{1}{2}\hat{\o}\hat{\tau}+\int_{C(T)} dt
(\hat{j}_{r}(t)\hat{e}_{r}(t)+\hat{j}_{\phi}(t)\hat{e}_{\phi}(t))
\l{3.29})\ee
and in $V_T (x,\vec{e})$ we must change $e_1$, $e_2$ on $e_{C,1}$, $e_{C,2}$,
where
\be
e_{C,1} =e_r \cos\phi -re_{\phi} \sin\phi,\;\;\;\;
e_{C,2}=e_r \sin\phi +re_{\phi} \cos\phi.
\l{3.30}\ee
The transformation looks quite classically but it
can not be derived from naive coordinate transformation of initial path
integral for amplitude: transformed representation for $R(E)$ can not be
written in the product form $AA^*$ of two functional integrals, i.e.
has not the factorization property because of the mixing of various quantum
degrees of freedom when the transformation is performed.

We can introduce the motion in the phase space with Hamiltonian
$$H_{j}=\frac{1}{2}p^2 +\frac{l^2}{2r^2}+v(r)-j_{r}r-j_{\phi}\phi$$.
The  Dirac's  measure becomes four dimensional:
\ba
D^{(4)}M(r, \phi ,p,l) =\delta (E+\o -H_T (r,\phi ,p,l))
\prod_t dr(t)d\phi (t) dp(t) dl(t)\times
\n \\
\delta (\dot{r}-\frac{\partial H_j}{\partial p})
\delta (\dot{\phi}-\frac{\partial H_j}{\partial l})
\delta (\dot{p}+\frac{\partial H_j}{\partial r})
\delta (\dot{l}+\frac{\partial H_j}{\partial \phi})
\l{3.31}\ea
Note absence of the coefficient $\prod_t r^2(t)$ in this expression. $\bullet$

The above example allows to note following general property of considered
formalism:

{\it S9. Arbitrary transformation of measure DM is `host-free'.}

Indeed, the transformation of measure
\be
DM(x)=\prod_t dx(t) \d (\ddot{x}+v'(x)-j)
\l{b1}\ee
can be performed inserting
$$
1=\int \prod_t dy(t)\d (y-Y(x)).
$$
In result,
\be
DM(y)=\prod_t dy(t) \d (\ddot{y}+Y'(X)X''(y)\dot{y}^2 +Y'(X)v'(X)-Y'(X)j)
\l{b2}\ee
since
\be
Y'(X)X'(y) \equiv 1
\l{b3}\ee
considering $X=X(y)$ as the inverse to $Y(x)$ function. $\bullet$

\section{Topology properties}\0

Let us consider motion in the action-angle phase space. Corresponding
operator has the form:
\be
\h{K}=\frac{1}{2}\int_0^T dtdt' \Theta (t'-t)
(\hat{I}(t)\hat{e}_I (t')+\hat{\phi}(t)\hat{e}_{\phi} (t'))\equiv
\h{K}_I +\h{K}_{\p}.
\l{3.32}\ee
The result of integration using last $\d$-function is
\be
R(E)=2\pi \int^{\infty}_{0} dT  e^{-i\h{K}}\int^{2\pi}_{0}
\frac{d\phi_0}{\Omega (E)}e^{-iV_T (x_c ,e_c )},
\l{3.33}\ee
where
$$\O =\pa h(I_0)/\pa I_0$$
with $I_0=I_0(E)$ defined by algebraic equation:
$$E=h(I).$$
The classical trajectory
\be
x_c (t)=x_c (I_0(E)+I(t)-I(T), \phi_0 +\tilde{\Omega}t+\phi (t)),
\l{3.33'}\ee
where
$$\tilde{\Omega}=\f{1}{t}\int dt' g(t,t')\O (I_0 +I(t'))$$.
The interaction `potential' $V_T$ depends from
\be
e_c=e_{\p}\f{\pa x_c}{\pa I}- e_{I}\f{\pa x_c}{\pa \p}.
\l{3.33''}\ee

One can note that eq.(\ref{3.29}) contains unnecessary contributions.
Indeed, action of the operator
$$
\int^{T}_{0}dt dt' \Theta (t-t') \hat{e}_I (t)\hat{I}(t')
$$
on $\tilde{H}(x_c ;\tau )$, defined in (\ref{3.7}), leads to the time
integrals with zero integration range:
$$
\int^{T}_{0}dt \Theta (T-t) \Theta (t-T) =0.
$$
This simplification was used in (\r{3.32}) and (\r{3.33}).

One can easily compute action of the operator $\exp(-i\h{K})$ since $\h{K}$
is linear over $\h e_{\p}$, $\h e_{I}$. The result can be written in
the form:
\be
R(E)=2\pi \int^{\infty}_{0} dT \int^{2\pi}_{0}
\frac{d\phi_0}{\Omega (E)}:e^{-iV_T (x_c ,\h{e}_c/2i)}:,
\l{a1}\ee
where
\be
\h{e}_c=\h{j}_{\p}\f{\pa x_c}{\pa I}- \h{j}_{I}\f{\pa x_c}{\pa \p}
\l{a2}\ee
and
\be
\h{j}_{X}(t)=\int^T _0 dt' \th (t-t') \h{X}(t'),~~~X=\p ,I.
\l{a3}\ee
The colons in (\r{a1}) means 'normal product': in expansion over $\h{j}_{X}(t)$
the differential operators must stay to the left of functions.

Now we are ready to offer the important statement:

{\it S10. Each term of perturbation theory in the invariant subspace can be
represented as the total derivatives over one of the cotangent
manifolds coordinate.}

This statement directly follows from definition of perturbation generating
operator $\h{K}$ on the cotangent bundle (\r{3.27}) and of translationally
invariance of the cotangent manifold in the classical approximation.

By definition $V_T$ is the odd over $\h{e}_c$ local functional:
\be
V_T (x_c ,\h{e}_c )=2\int^T _0 \sum^{\infty}_{n=1} (\h{e}_c (t)/2i)^{2n+1}
v_n (x_c),
\l{a4}\ee
where $v_n (x_c)$ is some $function$ of $x_c$. Inserting (\r{a2}) we
find:
\be
:e^{-iV_T (x_c ,\h{e}_c )}:=\prod^{\infty}_{n=1}\prod^{2n+1}_{k=0}
:e^{-iV_{k,n}(\h{j},x_c)}:,
\l{a5}\ee
where
\be
V_{k,n}(\h{j},x_c)=\int^{T}_{0}dt (\h{j}_{\p}(t))^{2n-k+1}
(\h{j}_{I}(t))^{k}b_{k,n}(x_c).
\l{a6}\ee
Explicit form of the function $b_{k,n}(x_c)$ is not important.

Using definition (\r{a3}) it easy to find:
$$
\h{j}(t_1)b_{k,n}(x_c (t_2))=\Th (t_1 -t_2) \pa b_{k,n}(x_c)/\pa X_0
$$
since $x_c =x_c (X(t)+X_0)$, see (\r{3.33'}), or
\be
\h{j}_{X,1}b_2=\Th_{12}\pa_{X_0}b_2
\l{a7}\ee
since indices $(k,n)$ are not important.

Let as start consideration from the first term with $k=0$. Then expanding
$\h{V}_{0,n}$ we describe the angular quantum fluctuations only. Noting
that $\pa_{X_0}$ and $\h{j}$ commute we can consider lowest orders over
$\h{j}$. The typical term of this expansion is (omitting index $\p$)
\be
\h{j}_1 \h{j}_2 \cdots \h{j}_m b_1 b_2 \cdots b_m.
\l{a8}\ee
It is enough to show that this quantity is the total derivative over $\p_0$.
The number $m$ counts an order of perturbation, i.e. in $m$-th order we
have $(\h{V}_{0,n})^m$.

$m=1$. In this approximation we have, see (\r{a7}),
\be
\h{j}_1 b_1 = \Th_{11}\pa_0 b_1 =\pa_0 b_1 \neq 0.
\l{a9}\ee
Here the definition (\r{3.24a}) was used.

$m=2$. This order is less trivial:
\be
\h{j}_1 \h{j}_2 b_1 b_2 =\Th_{21} b^2_1 b_2 + b^1_1 b^1_2 +\Th_{12}b_1 b^2_2,
\l{a10}\ee
where
\be
b^n_i \equiv \pa^n b_i.
\l{a11}\ee
Deriving (\r{a10}) the first equality in (\r{3.24c}) was used. At first glance
(\r{a10}) is not the total derivative. But inserting
$$ 1=\Th_{12}+\Th_{21},$$
(see the second equality in (\r{3.24c})) we can symmetrize it:
\ba
\h{j}_1 \h{j}_2 b_1 b_2 =\Th_{21}( b^2_1 b_2 + b^1_1 b^1_2)+
\Th_{12}(b_1 b^2_2+b^1_1 b^1_2)=
\n \\
=\pa_0 (\Th_{21} b^1_1 b_2 +\Th_{12}b_1 b^1_2) \equiv
\n \\
\equiv \pa_0 (b^1_1 \rar b_2 ~+b^1_2 \rar b_1 )
\l{a12}\ea
since the explicit form of function $b$ is not important. So, the second
order term can be reduced to the total derivative also. Note, that the
contribution (\r{a12}) contains the sum of all permutations. This shows
the `time reversibility' of the constructed perturbation theory.

$m=3.$ In this order we have:
\be
\h{j}_1 \h{j}_2 \h{j}_3 b_1 b_2 b_3=\pa_0 \{\sum^3_{i\neq j\neq k =1}
(i^2 \rar j \rar k ~+i^1 \rar j^1 \rar k)\}
\l{a13}\ee
since
\be
\Th_{ij}\Th_{jk}\Th_{ki}=0,~~~\Th_{ij}\Th_{jk}\Th_{ik}=\Th_{ij}\Th_{jk}
\l{a14}\ee
as the natural generalization of (\r{3.24c}).

So, $m$-th order contribution is following total derivative:
\ba
\h{j}_1 \h{j}_2 \cdots \h{j}_m b_1 b_2 \cdots b_m =\pa_0 \{
\sum^m_{i_1\neq i_2\neq \cdots \neq i_m =1}(
i^m_1 \rar i_2 \rar i_3 \rar \cdots \rar i_m ~+
\n \\
+i^{m-1}_1 \rar i^1_2 \rar i_3 \rar \cdots \rar i_m ~+
i^{m-2}_1 \rar i^1_2 \rar i^1_3 \rar \cdots \rar i_m ~+ \cdots +
\n \\
+i^{1}_1 \rar i^1_2 \rar i^1_3 \rar \cdots \rar i^1_{m-1}\rar i_m)\}
\l{a15}\ea

Let us consider now expansion over $\h{V}_{k,m}$, $k\neq 0$. The
typical term in this case is
\be
\h{j}^1_1 \h{j}^1_2 \cdots \h{j}^1_l \h{j}^2_{l+1} \h{j}^2_{l+2} \cdots
\h{j}^2_m b_1 b_2 \cdots b_m ,~~~ 0<l<m,
\l{a16}\ee
where, for instance,
$$ \h{j}^1_k \equiv \h{j}_I (t_k),~~~\h{j}^2_k \equiv \h{j}_{\p} (t_k)$$
and
\be
\h{j}^i_1 b_2 =\Th_{12}\pa^i_0 b_2
\l{a17}\ee
instead of (\r{a7}).

$m=2$, $l=1$. We have in this case:
\ba
\h{j}^1_1 \h{j}^2_2 b_1 b_2 =
\Th_{21}(b_2 \pa^1_0 \pa^2_0 b_1 + (\pa^2_0 b_2)(\pa^1_0 \pa^2_0 b_1))+
\Th_{12}(b_1 \pa^1_0 \pa^2_0 b_2 + (\pa^2_0 b_2)(\pa^1_0 \pa^2_0 b_1))=
\n \\
=\pa^1_0 (\Th_{21} b_2 \pa^2_0 b_1 + \Th_{12}b_1 \pa^2_0 b_2)+
\pa^2_0 (\Th_{21} b_2 \pa^1_0 b_1 + \Th_{12}b_1 \pa^1_0 b_2).
\l{a18}\ea
Therefore, we have the total-derivative structure yet. This property is
conserved in arbitrary order over $m$ and $l$ since the time-ordered
structure does not depend from upper index of $\h{j}$, see (\r{a17}).
$\bullet$

We can conclude, contributions are defined by boundary values of classical
trajectory $x_c$ in the invariant subspace since the integration over $X_0$
is assumed, see (\r{3.33}), and since contributions are the total derivatives
over $X_0$. One can say that the contributions are defined by topology
(boundary) properties of invariant subspace. For instance, the new
phenomena for quantum theories follows from $S10$:

{\it S11. The quantum fluctuations of angular variables are canceled if the
classical motion is periodic.}

If $x_c$ is the periodic function:
\be
x_c (I_0(E)+I(t)-I(T), (\phi_0 +2\pi ) +\tilde{\Omega}t+\phi (t))=
x_c (I_0(E)+I(t)-I(T), \phi_0 +\tilde{\Omega}t+\phi (t)).
\l{3.37}
\ee
this statement is elementary consequence of $S10$ and is the result of
averaging over $\p_0$, see eq.(\r{3.33}). $\bullet$

This cancelation mechanism can be used for the path-integral explanation
of quantum-mechanical systems integrability phenomena. The quantum problem
can be quasiclassical over the part of the degrees of freedom and quantum
over another ones. The transformation to the action-angle variables maps
the $N$-dimensional Lagrange problem on the $2N$-dimensional phase-space
torus. If the winding number on this hypertorus is a constant (i.e. the
topological charge is conserved) one can expect the same cancellations.

In the classical mechanics following approximated method of
calculations is used \C{arn}. The canonical equations of motion:
\be
\dot{I}=a(I,\phi),~~~\dot{\phi}=b(I,\phi)
\l{i'}
\ee
are changed by the averaged equations:
\be
\dot{J}=\frac{1}{2\pi}\int^{2\pi}_{0} d\phi a(J,\phi),~~~
\dot{\phi}=b(J,\phi),
\l{ii}
\ee
This is possible if the periodic oscillations can be extracted
from the systematic evolution.

In our case
\be
a(I,\phi)=j\partial x_c /\partial \phi,~~~
b(I,\phi)=\Omega (I)- j\partial x_c /\partial I.
\l{iii}
\ee
Inserting this definitions into (\ref{ii}) we find evidently wrong
result since in this approximation the problem looks like pure quasiclassical
for the case of periodic motion:
\be
\dot{J}=0,~~~\dot{\phi}=\Omega (J).
\l{iv}
\ee
This shows that the procedure of extraction of the periodic oscillations
from the systematic evolution is not trivial for quantum theories and this
method should be used carefully in the quantum theories.
(This approximation of dynamics is `good' on the time intervals
$\sim 1/|a|$ \C{arn}, i.e. for higher energy levels.)

\section{Conclusion}\0

Some remarks would be useful in conclusion.

--{\it Splitting of quantum excitations.}\\
In result of described splitting $j \rightarrow (j_{\th}, j_h )$ we
obtain a possibility to count the quantum excitations of each classical
degree of freedom independently. It is the truly Hamilton's description.
It allows distinguish `radial' and `angular' quantum excitations, i.e. allows
to consider the quantum fluctuations of bundles parameters and coordinates
on it independently.

--{\it Cancelation of quantum corrections.}\\
Investigation of integrable quantum-mechanical problems (of the Poschle-
Teller model, of the rigid rotator model) allows conclude that the reason of
total integrability is cancelation of quantum corrections. In this
sense the totally integrable quantum systems are trivial, quasiclassical by
theirs nature.

Note, the `level' of integrability of the problem
shifts position of singularities over the interaction constants. Indeed,
for the nonintegrable case the singularities are located at the origin, e.g.
\C{bw}. In the semi-integrable case the singularities at origin
are canceled \C{ush} and the main (rightist) singularities are located at
the finite negative values. And, at the end, the singularities of the
integrable systems are located infinitely far from the origin.

It is known that the case of totally integrable systems is very rear in the
Nature. But our secondary result is the observation that a quantum
problem can be integrable over part of the degrees of freedom. This fact
would have important consequences in the field theory.

--{\it Mapping on the cotangent bundle solves the nonlinear waves
quantization problem.}\\
The canonical transformation to the inverse scattering problems data for
sin-Gordon model is known (e.g. \C{kor}). Half of them
(solitons momenta) are the parameters of bundle and others (solitons
coordinate) are coordinates on the bundle. The problem of solitons
quantization would be solved by mapping of functional measure on the bundle
(see also \C{kor}). In result we get to $2N$-dimensional quantum-
mechanical problem quantizing $N$-soliton configuration and all above
offered statements becomes applicable for this field-theoretical problem
also.

In our terms the quantum sin-Gordon model is totally integrable
since the topological charge is conserved, i.e. the winding number on
the compact cotangent bundle is a constant. In result the quantum
corrections are canceled and sin-Gordon model is pure quasiclassical.
This fact for sin-Gordon model was noted firstly in \C{dash}). Our prove
of this result will be published later \C{wig, sin}.

\vspace{0.2in}
{\Large \bf Acknowledgement}

I would like to thank A.Ushveridze for very helpful discussions of
the semi-integrability phenomena. The work was supported in part by
the Georgian Academy of Sciences.


\newpage
\normalsize
\begin{thebibliography}{99}

\bibitem{gold}J.Goldstone and R.Jackiw, {\it Phys.Rev., \bf D11}, 1486
(1975);\\
R.Rajaraman, {\it Solitons and Instantons} (North-Holland Publ.
Comp., Amsterdam, New York, Oxford, 1982)
\bibitem{kor}L.D.Faddeev, {\it Solitons}, 363 (Ed. by R.K.Bullough and
P.J.Caydry, Springer-Verl., Heidelberg, New York, 1980);\\
V.E.Korepin anf L.D.Faddeev, {\it Theor. Math. Phys., \bf 25}, 147 (1975)
\bibitem{mar}M.S.Marinov, {\it Phys.Rep., \bf 60}, 1 (1980)
\bibitem{yad}J.Manjavidze, {\it Sov.Nucl.Phys., \bf 45}, 442 (1987)
\bibitem{smale}S.Smale, {\it Inv.Math., \bf 10:4}, 305 (1970), {\it ibid.,
\bf 11:1}, 45 (1970)
\bibitem{dow}J.S.Dowker, {\it Ann.Phys.(N.Y.), \bf 62}, 361 (1971)
\bibitem{untr}J.Manjavidze, {\it The Unitary Transformation of the Path-
Integral Measure} (Preprint, quant-ph/9507003)
\bibitem{ang}J.Manjavidze, {\it On the Canselation of Quantum-Mechanical
Corrections at the Periodic Motion} (Preprint, hep-ph/9509202)
\bibitem{fok}V.Fock, {\it Vestnik LGU, \bf 16}, 442 (1959)
\bibitem{mill}R.Mills, {\it Propagators of Many-Particles Systems},
(Gordon \& Breach, 1969)
\bibitem{arn}V.I.Arnold, {\it Mathematical methods of Classical
Mechanics}, (Springer Verlag, New York, 1978)
\bibitem{bw}C.M.Bender and T.T.Wu, {\it Phys.Rev., \bf D7}, 1620 (1973),\\
G.t'Hooft, {\it  The Whys of Subnuclear Physics}, ed.by Zichichi,
(Plenum, New York \& London, 1977)
\bibitem{ush}A.G.Ushveridze, {\it Particles and Nuclei, \bf 20}, 1185 (1989)
\bibitem{dash}R.Dashen, B.Hasslacher and A.Neveu, {\it Phys.Rev., \bf
D10}, 3424 (1978)
\bibitem{wig}J.Manjavidze, {\it Talk at Conference `Standard Model and
Beyond'}, ed.by J.Chkareuly, (Tbilisi, 1996)
\bibitem{sin}J.Manjavidze, {\it Path-Integral Description of sin-Gordon
Model on the Cotangent Bundle}, (Preprint, IP GAS-4-97, Tbilisi, 1997)

\end{thebibliography}
\end{document}